\documentclass[conference]{IEEEtran}
\IEEEoverridecommandlockouts
\usepackage{amsmath,amssymb,amsfonts}
\usepackage{algorithmic}
\usepackage{graphicx}
\usepackage{textcomp}
\usepackage{xcolor}
\usepackage{orcidlink}
\usepackage[english]{babel}
\usepackage[autostyle, english=american]{csquotes} 
\usepackage{cite}
\usepackage{xurl}
\usepackage{hyperref}

\begin{document}

\title{Publish Your Threat Models!\\
\small\textit{The benefits far outweigh the dangers}}

\author{\IEEEauthorblockN{Loren Kohnfelder, \orcidlink{0009-0001-1872-9399}}
\IEEEauthorblockA{Kalaheo, USA \\
loren.kohnfelder@gmail.com}
\and
\IEEEauthorblockN{Adam Shostack, \orcidlink{0000-0001-6837-5165}}
\IEEEauthorblockA{\textit{Shostack + Associates}\\
Seattle, USA \\
adam@shostack.org}
}

\maketitle

\begin{abstract}
Threat modeling has long guided software development work, and we consider how Public Threat Models (PTM) can convey useful security information to others. We list some early adopter precedents, explain the many benefits, address potential objections, and cite regulatory drivers. Internal threat models may not be directly suitable for disclosure so we provide guidance for redaction and review, as well as when to update models (published or not). In a concluding call to action, we encourage the technology community to openly share their PTMs so the security properties of each component are known up and down the supply chain. Technology providers proud of their security efforts can show their work for competitive advantage, and customers can ask for and evaluate PTMs rather than be told \enquote{it's secure} but little more. Many great products already have fine threat models, and turning those into PTMs is a relatively minor task, so we argue this should (and easily could) become the new norm.

\end{abstract}


\section{Introduction}
Structured, in-depth threat modeling techniques have been used for over 25 years to guide secure development of technical systems \cite{kohnfeldergarg1999}. Originally conceived as a tool for developers, we now realize they are also a powerful encapsulation of a product's security properties. Note that we are using the term \enquote{published} in its traditional sense, rather than the academic sense. Thus, this Position Paper centers on dissemination of information,  not on threat modeling. We anticipate future work could test some subset of these claims, perhaps comparing PTM to \enquote{privacy nutrition labels} or disclosures about product security. To the best of our knowledge, this is the first formal exploration of \textit{publishing} threat models, an extension of Saltzer and Schroeder's Principle of Open Design. This is a well-explored strategy for source code, and increasingly common with Software Bills of Materials (SBOMs). While focusing on software because it is the center of our own experience — and medical scenarios in particular as especially promising for early adoption — everything herein applies to digital hardware and perhaps an even broader scope of technology products.

Seeing great potential value to customers and the larger software community, we urge software makers to consider publishing threat models — if necessary, starting with creating them — we expect that it is the single most effective and accessible action to promote better customer and end-user understanding of security. 
Throughout this paper we will use the term \enquote{customer} to mean the people who are integrating a completed digital product, or possibly evaluating doing so.
Publishing a work product already done is a remarkably low-cost, relatively low-effort, and quickly achievable step that delivers an unparalleled return on investment for security communication. If no threat model exists, now is the best time to get started\footnote{See the \enquote{What is a Threat model?} Section for the justification of this assertion.}. We are well aware that some will have immediate strong responses, such as \enquote{this is a roadmap for attackers.} We address those concerns in the Frequently Raised Objections Section.

PTM precedents do exist: we mention several good examples to illustrate the benefits of publishing, as one of us has, repeatedly, to good effect\footnote{Zero-Knowledge Systems released  a set of \enquote{security analyses} white papers for our products, for example \enquote{Freedom 2.1 Security Issues and Analysis.} This 2001 document is not structured as a modern threat model, but explains the security choices we made in ways that meet the criteria presented in this essay \cite{backgoldbergshostack2001}. Similarly, we included a threat model for the Microsoft Threat Modeling Tool (v3) as an example within the tool.}. PTM aligns with a pillar of the \enquote{Secure by Design} guidance from many cybersecurity regulators: \enquote{embracing radical transparency and accountability.} \cite{ncsc2023} 

The recently enacted EU Cyber Resilience Act (CRA) requires many technology products to \enquote{include the cybersecurity risk assessment referred to in paragraph 3 of this Article in the technical documentation} (see Article 13(4)) \cite{eu2024cyberresilience}. Since threat models are an essential part of such a risk assessment, this is tantamount to requiring PTM (effective late 2027). Australia's Secure by Demand guidance for operational technology (OT) calls for a \enquote{full and detailed threat model} as a procurement requirement \cite{australiancyber2025}.

Without access to a threat model, for closed-source products we can only assume all threats are anticipated and fully addressed. Realistically, relying on such an assumption strains credulity, yet independent security assessment is difficult and expensive, PTMs offer a new option which buyers may start to demand. And even with the code in full view, deriving a complete threat model is a major undertaking, if it is even possible. Some choices may not be recorded, and some aspects (e.g. operations procedures and policies) are not in the code. Moreover, guessing about security is a risky game, and misunderstanding about responsibilities at just such boundaries is a common point of vulnerability.

We expect products with excellent threat models and significant investments in security work to lead the pack establishing PTMs as a new standard practice. And to the extent this puts pressure on competitors it serves as just the right kind of incentive for a healthy and more secure marketplace. 

\section{What is a threat model?}

A threat model document is the primary engineering record of the security analysis of a software or digital hardware design. It can take many forms, including an analysis document, a security operations guide, or as bug bounty rules (expressing in-scope threats). A PTM may be a lightly edited version of an internal document or a new document created to serve some outsiders. It may or may not be labeled with the words \enquote{threat model}.

A threat model document for a given product anticipates what could go wrong, how potential threats are addressed, and expresses overall security of the design. It can also show the limits of security aspirations of the designers (e.g., physical access, insider attack by administrators). Designers gain security insights from threat modeling that reveals how well the design handles many potential threats, indicating security vulnerabilities that need shoring up. When the design and its threat model show acceptable threat mitigation across the board, this is good evidence that security goals have been achieved.

It is important to note that threat models are important for IT systems\cite{yurcik2005}, and multi-system infrastructure as well. In fact, PTMs shine here because they allow integrators to understand the security of the whole from the models of the constituent parts.

In our view, designing or deploying digital technology without a threat model is \enquote{flying blind} with respect to risk, and only as PTMs become the norm will the fact and prevalence of such an omission become evident. Security vulnerabilities, even very severe and obvious ones, can go undetected for years because no attacker has for whatever reason bothered to look or act yet. Honest people often overlook these because their focus is on using the product as intended, and threat modeling is one of the best ways to surface just these issues. For all these reasons, a history of zero security incidents is hardly evidence that security is under control: think of threat modeling as a kind of due diligence. Readers interested in learning threat modeling should refer to either of the authors’ books \cite{Kohnfelder2021} \cite{Shostack2014} or the Threat Modeling Manifesto\cite{threatmodelingmanifesto}. 

\section{Historical perspective}
PTMs are important now because software development has evolved in ways unimaginable 25 years ago: with experience our scope has broadened, and because the software ecosystem has reached far greater scale and complexity.

Threat modeling was originally developed as a tool for software makers to systematically uncover security issues which had generally been underestimated if not largely ignored. However, modern software has a much deeper stack as evidenced by supply chain vulnerabilities, so it is important to think of security in terms of the broader scope rather than with each provider doing security work in their own silo. PTMs document security properties of the actual components and aid securing interfaces between them, enabling coordination across the larger community of customers, integrators, ops, testing, management, and end users.

Transparency, even to the level of distribution of source code, had a long history before the term \enquote{open source} was coined and focused attention on the value of openness \cite{dibona1999opensources}. More recently SBOMs have increased the transparency of components, and are scoped to exclude how those components interact. Demands for supply chain transparency and \enquote{third party risk management} are complemented by governments pushing for \enquote{Secure by Demand,} and all of these goals can be served by PTM.  

Disputes over openness have a very long history in cybersecurity, and many of those have related to vulnerabilities. The responsibility of a medical device maker to accept coordinated vulnerability disclosures is now well established, and many device makers run bug bounty programs, for example \cite{mckeon}. Despite those disputes, the value to technology creators, consumers and the public at large is now generally agreed. We expect a similar arc for PTM.

The recently released Cyber Hard Problems: Focused Steps Toward a Resilient Digital Future (The National Academies Press) illustrates how much software, and security in particular, has changed between now and then \cite{cyberhardproblems2025}. While earlier Cyber Hard Problems lists \cite{hardproblems1999} \cite{hardproblems2005} focused on OS and apps,  the themes of the latest Cyber Hard Problems derive from today's vastly more complex, highly connected, and interdependent world \cite[p. ~12]{cyberhardproblems2025}. PTMs directly address several of these critical hard problems with needed transparency and opportunities for coordination; specifically, Problem 1: Risk Assessment and Trust; Problem 3: System Composition; and Problem 4: Supply Chain, in addition to contributing context and understanding essential to tackling several others on the list as well.

There is a complementary stream of work emerging under the label of Software Understanding which we believe PTM can support by expressing explicit design intention. 

\section{Context matters}
Threat modeling depends on and explains context. The same component deployed to various systems will be exposed to very different sets of potential inputs, for example depending on if it is within a private network or on the public internet. Assumptions about available resources will vary, and these will determine how much compute demand constitutes denial of service. The consequences of information leakage are extremely context dependent: a transaction ID may be meaningless and also harmless, while a SSN could be used for identity fraud.

Beyond system context, the notion of potential harm depends on whose perspective is being considered. An example of a wirelessly connected medical device demonstrates this kind of context clearly, where a doctor prescribes the device to be implanted in a patient. Only healthcare professionals ever interface with the device for configuration and monitoring, yet if the communication protocol broadcasts sensitive personal information 24/7 that is a serious privacy compromise from the patient's perspective. A PTM for a software product allows its evaluation within each customer's particular context, which the provider can only speculatively consider.

\section{How secure is your system?}
What is the best way to answer the question of how secure is your system? Answering it is difficult because everything hinges on what the single word \enquote{secure} means, and if you ask a number of customers they probably will all give different answers. The technology provider's job is to somehow deliver a product that meets the requirements of most customers, which is precisely why PTMs are a great way to answer this very important question with substance and precision.

When a product has a PTM the provider is showing what threats were anticipated by its design, and explaining what protections are included to address those threats. Customers can decide if the level of security covers the threats they care about. If all the customer needs are covered and the defenses are satisfactory, then the product should be secure enough (assuming reasonable care in implementation).

If not, the PTM has provided enough detail to clarify where it falls short and to what degree, so a conversation about the difference is possible. To appreciate how powerful this is, consider how to have the conversation without a PTM. Both parties have some notion of what \enquote{secure} should mean but how do they compare those and home in on perhaps a minor discrepancy? For a large software product very few people, if any, would have the knowledge required to answer security questions authoritatively.

\section{Precedents}
We have not done a formal survey but have found several real world examples that serve as early evidence of the value of PTMs. While some of these are not labeled as threat models per se, by answering the Four Question Framework\footnote{Shostack's Four Question Framework is \enquote{What are we working on, what can go wrong, what are we going to do about it, did we do a good job?} See  \cite{4qwhitepaper}} they function as PTMs. Examples include:
\begin{itemize}
    \item \textbf{Zero-Knowledge Systems'} Freedom 2.1 Security Issues and Analysis \cite{backgoldbergshostack2001} Shostack, one of the authors of this Position Paper, was an author of this and related corporate white papers.
    \item \textbf{SecureDrop} \cite{securedrop2024}, explaining to their users how they think about the security needs of their users.
    \item \textbf{NIST} has created a generic threat model for genomic data applications for community review, gaining insights and validation from the responses. \cite{nistcswp35}
    \item \textbf{Meta} \cite{meta2025} blogged their \enquote{threat model for Private Processing} (\enquote{a confidential and secure environment} for AI), read whitepaper \cite{metawhitepaper2025}.
    \item \textbf{Apple and Google} developing and publishing \enquote{Risks and Mitigations} \cite{exposurenotification2020} for the COVID Exposure Notifications project, which interestingly mirrors the ZKS paper in covering how security protects privacy. NOTE: The linked design doc is now archived here \cite{applegoogle}.
    \item The open source \textbf{cURL} project's cURL Threat Model Report \& Fix Review - 2022 \cite{trailofbits2022}.
    \item \textbf{Kubernetes Threat Model} – SIG-Security External Audit 2019 \cite{kubernetes2021}.
    \item \textbf{HashiCorp Vault Secrets Operator Threat Model} – GitHub \cite{vaultssecret2023}.
    \item \textbf{Kata Containers Threat Model} – GitHub \cite{kata2024}.
    \item \textbf{Fediverse End to End Encryption Spec}, Public Key Directory Server, GitHub, threat model for the Fediverse Key Transparency specification \cite{publickey2025}.
    \item \textbf{FIDO Alliance Security Reference}, 27 Feb 2018 \cite{fido2018}.
    \item \textbf{A Threat Model of High-Power Electric Vehicle Charging Infrastructure} \cite{highpower2022} is a generic framework of threats (without mitigations) but serves as a hardware example that makers can readily build on. 
    \item For even more examples see Padmos' extensive list of threat models (some may not have all the elements we want, but we applaud partial security disclosures as better than nothing) \cite{additionalexamples}.
\end{itemize}

Readers may find good ideas for approaching and writing up their own threat models from perusing these examples which cover a diverse sampling of technology components and threat model forms and styles.

Many of these documents share the understanding that their security properties are explicitly in tension with other goals. Using Zero-Knowledge Systems as an example, we were aware of speed-security tradeoffs, that security was improved by adding users, and that users wanted speed.

Several of these are not labeled threat models, but we care less about the label than content. We found the publication simplified debate with security experts by clarifying intent. Designing the Exposure Notifications protocol during the COVID lockdown was perhaps the most contentious software project in recent memory; regardless of how well they succeeded, with a public \enquote{Risks and Mitigations FAQ} at least the debate was informed.

In each of these cases, the publication of a threat model improved the conversation about the tradeoffs being made across requirements, across disciplines and knowledge domains, and across technical implementation choices. For example, many objections to COVID exposure were based on imagined designs, or a fear that the design had not been analyzed. Having a publicly available document addressed those concerns, which otherwise would have lingered. Some may claim that their systems involve no tradeoffs that customers need to know about. We disagree: if the system has security implications then there are almost certainly tradeoffs being made.

\section{Benefits of PTMs}
In this section we list major benefits that public threat models afford to various parties as a substantive description of the security properties of a technology product. 

\subsection{Benefits to providers}
The value of threat models to guide secure system design and implementation is well known, so here we focus on value to the provider when they make their threat models public. Having already invested effort in a threat model, releasing a PTM lets the provider share their own success story, demonstrating the thoroughness and quality of the work. Until PTMs become a routine part of technical documentation, early adopters will distinguish themselves having PTMs compared to the competition who can only say, \enquote{trust us}.

Further benefits from PTMs accrue as they are reviewed by analysts and customers who should appreciate the transparency. Some will have questions and suggestions, with the PTM giving them first a good understanding necessary to foster a productive discussion. Should a customer raise a new threat, the provider can consider ways of mitigating it or perhaps advise how it could be handled externally.  

PTMs are potentially a powerful sales tool because great threat models substantiate security claims by making it concrete. Publishing the design of a system has served business interests including Zero-Knowledge, Apple, Google, Meta, by showing off their privacy investments, explaining the tradeoffs they have made that might otherwise appear insecure. At Zero-Knowledge, publishing our threat model allowed us to avoid some criticism that otherwise would have been tedious to address.

Many providers do not currently have explicit, integrated, or documented threat models. While we would not mandate that they create threat models, it is reasonable to wonder if their security protections are cogent or coherent. It is hard to imagine that those products would not be improved by the work to document their threat model. Possibly one person understands the product entirely and has managed to secure it by extraordinary effort, but should not the knowledge be shared for greatest effect and their eventual departure? To the extent that work is expensive, it is likely expensive because their models are (at least) baroque or more likely, inconsistent. There is a reasonable question of whether the investment in threat modeling is worthwhile. We refer those who would hesitate to the legally binding promise in their privacy policy that \enquote{your privacy and security are important to us.} With less snark, in this scenario the cost is certainly higher than if a threat model already exists, and we think the effort to document the threat model and improve the system's architecture are both good investments.

We should ask if a provider gets these benefits from providing the same documents under NDA. They can obtain some of them, to a lesser degree, or at a higher cost, but it certainly is better than nothing. For example, the assurance that comes from wide review and public discussion will be missing, as well as making it burdensome for prospective customers to evaluate.

Today, technology providers are increasingly burdened by what their customers call `third party risk management' (TPRM) processes. TPRM usually takes the form of either contractual clauses or a \enquote*{questionnaire.} Both types are expensive, and personal experience leads us to believe that neither side is happy. PTM could be an innovation which changes the burden or delivers more value, and certainly informs discussions of risk management between parties.

\subsection{Benefits to customers}
Customers must choose from an ever wider range of software products, and for modern interconnected systems careful evaluation, integration, and deployment are especially important. These tasks may have different steps, but each involves a security analysis which can be arduous, requires technical skills and judgment, and involves \enquote{rebuilding} knowledge that might well be encoded in a threat model done by the designers. To the extent that re-doing the work is a validation step, it may be useful. The usefulness may be that simply identifying that the PTM is wrong is indicative of quality issues that the analyst cares about, or that identifying errors can get them fixed. That analysis depends on an understanding of how the system is organized.

However, for proprietary technology there probably will not be sufficient information available to threat model as quickly or completely as the provider could — and should have already. Even with full open source, for larger codebases the effort involved creating an in-depth threat model is extensive (not to mention requiring great expertise) and still would require guessing at what threats the design anticipated and how they addressed them. This may tie to the failure of the \enquote{many eyes} argument that security analysis is easier with open source software; the analyst needs to first construct a threat model from code, then check it, then find bugs \cite{pell2004}.

With a PTM, a customer can quickly understand and assess the risk profile of a system and the intentions of its designers. To the extent that it does not match their needs, they can assess if “compensating controls” can be \enquote{bolted on} -- (often this is a risky move contrary to Secure by Design, but full knowledge of the threat model should help considerably.) A PTM that omits obvious threats may be a red flag indicating that the provider under-invested in security, potentially questions their competency, or it may be due to misalignment between the producer and 
customer.

Beyond assessment, a PTM can be a road map for interfacing with the system or component. Knowledge of mitigations, or their absence, enables informed decisions by integrators. For example, deciding where input validation is needed, or not if handled by the third party component according to its PTM. Alternatively, for a security-critical application additional mitigations using different techniques might be layered on for defense in depth. Conversely, where performance is key, redundant mitigations (deemed necessary for a black box component without a PTM) can be confidently removed.

Some prospective customers may see a PTM as an invitation to debate the architecture. Vendor openness to this will vary, and often by the time a product is offered to customers, the security decisions are set. For those vendors, the value of publishing a threat model is clarity in procurement, purchase, and deployment.

Commercial software covers a spectrum from highly customized to mass scale releases, and published threat models are accordingly used differently. While customizable software threat models may be used as a basis for refining security requirements and trade offs, for off-the-shelf software they offer clarity for procurement decision making and deployment. When customers see gaps that will not be addressed by the provider, the threat model does provide a clear map showing which threats they would need to defend (e.g. through layered security).

\subsection{Benefits to others}
PTMs benefit industry analysts performing product assessments, as well as security researchers investigating vulnerabilities or testing effectiveness of fixes.

Finally, we look forward to a future where PTMs are widespread when they foster effective communication and serve as examples across software communities. Just as open source promotes sharing and learning at the source code level, PTMs will unlock great threat modeling and help establish broadly accepted practices when we all show our work.

\subsection{Benefits to end users}
Sometimes, the person using the technology is not the customer with the traditional connotations of choosing or paying for a product. For example, a doctor prescribes a device to their patient, and an insurer pays for it. In the medical device case, the dangers of the product are nominally included in \enquote{The Label,} a term of art that seems to require 9 point text in a light gray font enumerating anything that has ever gone wrong within a 100 foot radius of the product. Would it be helpful to add security information, or would that just exacerbate current issues with labels? For many of the security issues, it is easy to say the right end user solution is \enquote{don't enable that.} Also, many users of medical devices may need assistive technology, or even assistance using the device in ways that are in tension with traditional notions of security. For example, a person who has lost a hand cannot use a fingerprint reader, and may have a lot of trouble typing or otherwise entering a password.

\subsection{Health security: example of benefits}
This section discusses how threat models help the relationship of medical device makers (MDM) to Healthcare Delivery Organizations (HDO), including hospitals, and argues that publishing threat models saves time and energy without sacrificing anything. Before we get there, there are other customers and users of medical devices which we consider briefly. As mentioned above, \enquote{a doctor prescribes a device to their patient, and an insurer pays for it.} Most patients are not technically sophisticated consumers, and some may be alarmed by the mention of a \enquote{threat model.} 

Medical device makers selling in the US already provide threat models to regulators as part of the review and approval process, so they are already doing most of the work to produce a PTM \cite{rabbitholes2024}. The value of a PTM to hospitals, labs or other sophisticated customers of medical devices is evident in aid of ensuring customer safety and privacy throughout the healthcare delivery process. Given this alignment, we expect that the medical device industry may be among the first to adopt the practice of PTM given minimal additional cost and great potential benefit.

The model \textit{Device Security Lifecycle for Healthcare Delivery Organizations} in \cite{wirth2024medical} consists of: Pre-procurement, Procurement, Deployment, Operations, and Decommissioning. PTMs can drive down the cost for both vendor and customer while increasing security in each phase. They write, \enquote{The HDO lifecycle is centered around the actual device and its target integration ecosystem, supported by documentation and artifacts provided by the manufacturer, and in the form of risk mitigations during the HDO's deployment and operations phases.} Threat models could be a part of that, and if as accessible as other documentation, then work will proceed more smoothly, saving time and money.

In the Pre-procurement phase, the HDO considers \enquote{security expectations and requirements as a buying criterion.} PTMs inform the expectations and requirements, especially if devices within a specific class have different threat models. For example, one infusion pump only sends its logs to the manufacturer’s cloud, another requires configuration for either storage on the device, on prem, or on cloud. These choices might well jump out in a quick glance at a diagram. Here, a good threat model may actually disqualify or disadvantage competitors by making an innovative security difference clear. A threat model only available to customers or under NDA is unavailable at this phase so this benefit requires a PTM.

At the Procurement phase, a PTM supports elements of the Health Sector Council's Model Contract Language for Medtech Cybersecurity (MC2)\cite{HSCC_MC2_2022}. The MC2's \textit{Maturity Pillar} includes \textit{Security Development Lifecycle} and \textit{Supplier Transparency} \cite[page 8]{HSCC_MC2_2022}. Each of these is clearly buttressed or demonstrated by a PTM. For contracting purposes, Clause 37 (Secure Design) may require a threat model to exist. The language may be read to require an \enquote*{Assessment,} which may be slightly different. Similarly, there is a \textit{Product Design Maturity Pillar} (\textit{Secure by Default} and \textit{Standard Security Controls}). Perhaps surprisingly, these may not be so clearly related to a PTM, but they are directionally aligned.

In the Deployment and Operations phases, easy access to the threat model, especially the Global System View and Security Use Case Views the FDA recommends for Pre-Market submissions enables deployment planning and operations to determine if their deployment and operations match the documented expectations of the MDM.

Finally, in the Decommissioning phase, the threat model is a guide to where confidential patient or operational information will need to be deleted, and perhaps more factors. While this should be covered in the decommissioning guidance document, the threat model provides context and could help remediate in the case of lax decommissioning procedures.

Many of these functions are already handled to some degree. A PTM provides a holistic perspective that can enhance those functions, reducing confusion, re-using work that is already being done, possibly replacing other documents (again reducing MDM costs).

\section{Assessing PTM}
\subsection{Frequently Asked Questions about PTMs}
We anticipate that more than a few readers will consider this is a pollyannaish notion, but for many technology offerings (including open, proprietary, and custom software, sold or given away) it is worthy of consideration when security matters. First, it is important to clarify our position that PTM is a generally valuable practice to aspire to, from possible misunderstandings. We are \textit{not} suggesting that all software products must have a PTM, nor that lacking one implies security weakness, but we do think that doing so would be significantly beneficial. Choosing to publish threat model documents is a transparency choice, not unlike open source, and it offers closed source software providers an option to disclose security details for customer assurance within that model.

Modern software systems are often large, complex assemblies involving containers (themselves assembled from components), cloud services and more. The complexity of these assemblies creates a 
need for robust sharing of substantive security information between providers, customers, and a host of partners. Some providers may feel that they can take care of security on their own and find PTM at cross purposes with this stance, yet we would like to point out that rarely can this strategy of isolation succeed for components that interact with a diverse set of applications. For example, consider `auditing the logs' which is clearly something the customer must do (they know their own data, plus data privacy). To do this effectively they need to know about the component producing the logs in order to identify potential attacks.

There is a question of \enquote{do we need models to be standardized before publication can be beneficial?} Standardizing the form and methodology for threat modeling (TM) could ease and unify the approach, but would add the cost of conversion and delay near term publication.  Our goal here is focused on the low hanging fruit of disclosing all existing TM, using the Four Question Framework as the broadest possible definition of a PTM. We believe that premature optimization could lessen the value we might get from varied threat models. With more disclosure there is an opportunity for the community to learn what others are doing and how to best present PTM; in this way transparency may enable standardization that has long eluded TM practice. By providing a familiar structure, standardization also eases review and analysis, whether it be algorithmic, LLM, or human.

Today, the forms that PTMs take vary widely. SecureDrop\cite{securedrop2024} uses a form of \enquote{actors, assumptions, and assets,} with an unusual extension of \enquote*{actors} to include both authorized users and adversaries.  Their \enquote*{actors} also includes systems, which others list as assets. We point this out not to critique, but to spotlight the diversity that PTMs take today.  (Interested readers can see more at Threat Model Thursday \cite{threatmodelthursday}.) We expect that seeing more PTMs, we will learn much about their prevalence, diversity, and commonalities, which is useful information for anyone seeking to standardize them. For better or worse, we suggest taking the lack of standardization as liberty to publish your threat model in any suitable format. So long as it comprehensively treats recognized threats and answers in the Four Question Framework, it will deliver the benefits we describe.

We anticipate that some will want to cut from the full model those threats that are completely resolved in the product, believing those details to be irrelevant to other parties. We strongly recommend against such simplification (other than redaction for specific reasons as explained later) because the complete threat model is far more valuable. Specifically, threats considered fully mitigated internally and hence no concern of customers should be included in the model. That way others will know the full set of threats the developers anticipated and how they are mitigated — without the details others are left to wonder. In contrast, threats fully eliminated by re-designing the system would be a choice for the publisher. It shows good diligence, and if the threat is moot then the entire issue can and should be omitted. For example, if a system had a network interface that has been removed, then all the issues with that network interface are now eliminated. In other words, a PTM is useful to the degree that it expresses these details. The absurd limit of \enquote{publishing a threat model} would be a single sentence stating, \enquote{All threats discovered in threat modeling have been fully mitigated.} (As an aside, such a statement might be seen by an activist FTC as a representation that there are no security issues in the product. Good luck ensuring it’s not a deceptive trade practice.)

\subsection{Frequently Raised Objections}
PTMs conflict with security by obscurity in a good way, showing how concerns about disclosing the threat model may point to areas that need improvement for safe disclosure. When a PTM would aid attackers it is probably due to a real vulnerability, and as such is best addressed by fixing or redesign rather than concealment. It could also be that the threat model exposes implicit assumptions that lead to an attacker seeking to violate them. Whether this is a vulnerability or an important red flag to be urgently addressed, is a matter of perspective. We believe that publication will help potential customers make better decisions. While it is tempting to paper over such issues in the short term, PTM at least offers an opportunity for good faith resolution of these problems before vulnerable systems are deployed and widespread harm ensues.

Any threat model should be safe to disclose once scrubbed for private or internal information as we describe following. Threat models describe how you have secured your product, not how to break it. If disclosure feels risky, consider if that is due to fundamental, unaddressed vulnerabilities and if there are not better defenses than hiding them. We have done a lot of threat modeling and have only seen such a conundrum when threat modeling after a product has shipped due to back-compatibility demands. When maintaining compatibility comes at the expense of latent vulnerability, it is important that customers be aware of such security trade-offs so they can prioritize upgrading to a newer, more secure version: so even in this case, threat model sharing is beneficial.

A good threat model surfaces these problems so they can be addressed, not hidden. Therefore, disclosing a robust threat model demonstrates confidence and invites scrutiny to improve security, not weaken it. A good example of this is that the FIDO Alliance protocol defense against monkey-in-the-middle is incompatible with TLS proxies \footnote{FIDO Security Reference T-3.1.1: \enquote{Channel Binding may indicate a compromised channel even in the absence of an attack.}}. Customers need to choose between two security goals — requiring channel binding means the system breaks if a proxy is in use, but channel binding has security value — and a PTM can precisely lay out the trade off.

Many see security by obscurity as an important defense tool. If it is being relied upon, PTM probably is not viable because it exposes that strategy, allowing attackers to find and focus on those places, or by inviting customers to object to the choice. Instead, we favor the Open Design\cite{saltzerschroeder1973} security principle, which holds that security should never depend on hiding the design, aligns with PTM because that is the abstraction layer where threat modeling happens. So if it is visible at the level of a threat model, it is probably also difficult to change, and if there was an easy fix there would be no reason to conceal. While creators will not like it, this is security relevant information, and prospective customers probably would object vehemently should they learn about it the hard way.

Perhaps a secret military facility is the \textit{ne plus ultra} of where obscurity (or even secrecy) can be effectively deployed. If that is so then the limits of the value of secrecy may be illustrated by the Natanz nuclear facility. The Iranians presumably worked hard to keep the operational details a state secret, which did not prevent the creation or apparently successful deployment of Stuxnet \footnote{We look forward to hearing from the elite hax0rs whose systems are better secured than Natanz.}. In other words, dedicated attackers create their own roadmaps.

As PTMs become an accepted norm, the products that rely on security by obscurity will be clearly identified as such, giving customers better information in the market. 

Some may accept the value of PTM but suggest that only covered threats need be public, without going into detail about mitigations. However, doing so amounts to partitioning by threat and simply stating without basis that the threat is covered: \enquote{Information leak between server and client: trust us}, etc. Mitigation details are useful to others assessing the PTM and also add depth to the particulars of the corresponding threat. For example, protecting a publicly exposed API by doing input validation addresses syntactic abuse attacks but clearly ignores attacks playing with semantics. Furthermore, recalling the Four Question Framework, without a description of mitigations it is impossible to assess \enquote{Did we do a good job?}

\subsection{Forthcoming Regulatory Changes}
Whatever you think of our arguments, the European Union is changing the equation. The Cybersecurity Resilience Act, requires manufacturers to implement security by design and conduct risk assessments, and to document such as part of product documentation. For example, \enquote{At minimum, the product with digital elements shall be accompanied by...  any known or foreseeable circumstance, related to the use of the product with digital elements in accordance with its intended purpose or under conditions of reasonably foreseeable misuse, which may lead to significant cybersecurity risks.} (CRA Annex II, §5.) These requirements are magnified for \enquote{important} or \enquote{critical} products. The work to get a threat model \enquote{publication ready} will become more of a requirement, reducing the arguments against it, and raising questions of why extant threat models are not publicly available.

\section{Preparing to publish TMs}
When security matters and the benefits of publishing the TM for a product are compelling, we recommend a basic process for taking existing internal documents and making them public. The effort involved should be straightforward, and should be proportionate to the size and degree of formality of the internal TM. 

The range of methods and styles of threat modeling is so varied that no single process works for all, but we offer high level guidance that should be broadly adaptable. The basic steps, detailed following, are:
\begin{itemize}
    \item Collect copies of the original TM and supporting documents.
    \item Adjust level of detail and precision for the external audience.
    \item Cull internal documents or sections unrelated to the TM.
    \item Redact any trade secrets or confidential information.
    \item Add context where the TM depends on anything deleted.
    \item Review looking for any security implications if disclosed.
    \item Consider encoding in a standard form.
    \item Clean up editing as needed to publication standards.
    \item Final review and sign-offs for publication.
\end{itemize}

Here we are considering a wide range of potential issues for a PTM, but in practice few of these precautions will apply to most TMs because typically these are design-level treatments, not implementation details. This process largely targets identifying and remedying those cases where extraneous information leaks in. 

It is important to note that the effort of publishing may be rewarded by useful feedback \footnote{As noted above, there is also a chance of a pointless argument.}. Customers, analysts, security researchers, and more now have an opportunity to understand the threat model and if necessary raise issues. Ideally, the original, nominally private, threat model document can be published as-is. In some cases it may need improvements to serve a broader audience, which we think is a reasonable cost that serves customers.

This section is intentionally written in a directive voice to make application easier. For clarity, we focus on advice for a \enquote{usual} case without trying to foresee every variant. Technology providers know their product, customers, and threat landscape best and how to apply this guidance with judgment. 

Disclosing a threat model only makes sense where the benefits exceed the effort and risk. Disclosing any information about a technology product (documentation, release notes, and so on) potentially aids attackers in some small way, but doing so is common practice because it helps good people to a much greater degree.

\subsection{Collect documents, adjust detail level}
Start with the core TM document(s), add supporting documents referenced therein or important as a foundation. Where existing customer documentation stands in, refer to that instead. 

For large highly detailed TM, often a lower resolution summary will be most suitable for the PTM. Consider using an existing summary or reducing the level of detail, balanced against delivering the benefits as a PTM, perhaps entire documents are not needed for the core ideas. Where existing threat models exist that are factored to a different scope than the public-facing component or app, consider adjusting scope to match the customer perspective.

A threat model might be kept light on important details specifically because it is expected to become public.  A published threat model might say \enquote{Our trade secret X defends against Y,} and customers can decide if they feel that’s sufficient. We are aware of no case where redaction of sensitive internal details is not enough for a secure product to have a PTM. (If part of the threat model seems risky to disclose that is not easily rewritten to resolve, consider carefully if this is not a vulnerability that needs fixing if mentioning its existence is dangerous.) 

Even in the case of things like threat models for currency counterfeiting or nuclear weapons, where many details are kept secret, the designs are believed to be secure even if the details would be public. For example, those who print legal tender include a set of secret details that they can use to detect forgery, but that fact is known, even if the specific details are not. Thus the Easter eggs work as defense in depth. Similarly, the specialized paper makers will only sell each currency paper to a few customers. (The paper used for, say, the Pound, is different from the paper used for the Euro.) So even knowing all characteristics of the paper, getting it made is hard.

\subsection{Remove internal-only parts, add back needed context}
Next step: review the document(s) with an eye to it being published. This consists of not only spotting anything unsuitable to being public, but also as a reader without the background knowledge of an insider to fill in the blanks enough for them to follow.

\begin{itemize}
    \item Private information leakage: while threat model documents typically refer to design level abstractions, it is possible that some internal details inappropriate for disclosure are included. Usually the names of individuals and other administrivia should be redacted. Look for references to internal links to tickets or bug reports that will require explanation, or should be removed. We recognize there certainly are reasons that more threat model documents have rarely been published to date, and would like to address those concerns in light of the potential benefits described above. 
    \item Completeness: ensure that all relevant threats are addressed; omitting an unmitigated real threat is a good hint for attackers.
\end{itemize}

Where removals created holes, write an introduction and give context that would be either well-known to insiders, or covered in internal documents that the company wishes to keep proprietary. 

\subsection{Security implications of publishing}
At this stage the PTM is almost ready: material inappropriate for publication is out, any gaps created have been filled, and outsiders will be able to read and understand the document. As for any public disclosure, it may be worthwhile to consider how malicious actors might also read to their advantage. 
\begin{itemize}
    \item Omitted real threat: vulnerability to unanticipated threats is what threat modeling aims to fix proactively, and making public such a mistake is indeed a danger. However, we argue this is a serious failing worth addressing eventually, and the act of publication can serve to identify the issue ideally as part of a pre-publication review, or by public comment. 
    \item Redacted threat (known and mitigated): omitting a threat from the PTM that is in the TM is not an actual danger, but publicly it will look like one so there is potential reputational damage.
    \item Redacted mitigation: sharing how threats are mitigated in general demonstrates how secure the product is and should be an unmitigated benefit. Conceivably a provider might not want to disclose the \enquote{secret sauce} of a novel mitigation, but we would still advise noting it without details, such as, \enquote{Information disclosure prevented by our exclusive Magic Security Dust \cite{magicsecuritydust}}. 
    \item Omitted or insufficient mitigation: closing the door on potential threats is the other major goal of threat modeling, so making such a weakness public is also a danger. As above, this is a real problem and publishing potentially makes it easier to find and exploit.
    \item Helping the competition: there may be disreputable providers who copy another PTM as a short-cut. Not only is this thoroughly unethical, but also they may be shooting themselves in the foot doing so as they may be unwittingly susceptible to many dangers by not doing important security work. Any mismatch to the product may soon be identified by the fake PTM, as a possibly important corrective.
    \item Effectiveness: check that all threats are thoroughly mitigated; missing or ineffective mitigations are invitations to attack.
\end{itemize}

\subsection{Clean up, final review}
Once a draft is ready, it may need to be spruced up with graphics and branding, or you may find that a technical feel conveys something important. It may help to have a few customers review a draft.  In addition to the normal publication review, possibly executive and/or legal sign-off.

\subsection{Keep it up to date}
When a PTM is released is an ideal time to plan for updating it going forward. In general, consider updating a PTM, or for that matter any TM, when:
\begin{itemize}
    \item New features or architectural changes are planned: first internally to guide design and implementation, to be folded into the PTM in concert with release;
    \item Lost sales opportunities citing concerns about security;
    \item A security incident occurs: learn first, then update as needed;
    \item New tactics are illustrated by prominent security incidents, possibly with other products. (Ask, \enquote{Could this kind of attack work against us?})
    \item Tactics including the misuse of features become increasingly common because of scripting or widespread knowledge of their efficacy.
    \item New attack vectors become possible because of changes in the environment such as new OS features (e.g. Microsoft Recall starting to record messages that are otherwise transmitted and only stored in encrypted form.). 
    \item A vendor takes a dependency on the mis-use of that feature.
    \item User knowledge or skill proves insufficient to operate the product as hoped. (This may include software designed for power users that will now be deployed broadly, especially if the protection strategy depended on flawless end user decisions.)
    \item Periodic reviews (such as annual) of the PTM give assurance that it remains accurate, considering the slowly evolving threat landscape, customer or analyst feedback. 
    
\end{itemize}

You should also consider updating the model if customers or prospects routinely get stuck on either the presentation or reality of an element of the PTM.

\section{Call to action}
Threat models are the best methodology we have to specify the security properties of technology, explain what the design anticipates, and how protection is provided. Rather than making analysts, integrators, and users of products \enquote{reverse engineer} a threat model themselves, the maker helps everyone by curating updated PTMs for all stakeholder perspectives to aid understanding and ensure better security.

Building a secure modern system is only possible through understanding and cooperation across all components, operating in concert with the enterprise and users it serves. In other words, security is a team sport. PTMs enable rich two-way communication between providers and customers that is needed to keep up with the dynamic threat landscape. Hidden (or imagined) threat models inhibit just such understanding, forcing customers to guess about security, and also risk there being unaddressed threats on the provider side as well.

We encourage the technology community (including both commercial and open source providers) to take action and publish their threat models, sharing detailed views of security in the interest of long term synergy. If you make security claims, and have a threat model that you cannot publish because it is useful to attackers, perhaps your security claims are better characterized as puffery, overblown, or incomplete.

We encourage technology consumers to demand threat models from providers, to incorporate threat model evaluation into procurement and risk management programs including but not limited to the corporate function called \enquote{third party risk management.}

Once providers disclose PTM we look forward to seeing the diversity of work products, learning how they are used in practice, and eventually research measuring impacts and costs. We also hope to see TM tools that assist preparation for public release, and TM education to include producing PTM as well as utilizing those of dependencies in larger system analysis. Proposing the disclosure of countless threat models that exist behind NDA walls we can only speculate about how they will be used, but we have no doubt many more innovative uses will be discovered once the information is widely available.

PTMs allow makers to show off their excellent security work, meaningfully characterize what the product is designed to protect against and how it does that. They allow customers to make better informed decisions in purchasing and deploying systems. Stakeholders will know exactly what their part is in overall system security, have an implied commitment by the maker, and have actionable information for defense in depth or other enhancements. With growing adoption over time, customers might begin asking \enquote{Where is your threat model?} and products without one could perhaps become increasingly difficult to justify.

Or we just continue hiding valuable security information from customers and partners who must blindly trust that makers are diligently threat modeling behind the scenes — while worrying that perhaps they do not but somehow totally have security well in hand. What could possibly go wrong?

\section{Acknowledgments}
Zoe Braiterman, Matt Coles, Josiah Dykstra, Christopher Gates, Bill Hansen, Robert Hurlbut, Matin Mavaddat, Takaharu Ogasa, Bruce Schneier, Izar Tarandach, Petra Vukmirovic, and William Yurcik all provided helpful comments on drafts. Kris Yun provided helpful editorial support.

\bibliographystyle{IEEEtran}
\bibliography{references}

\end{document}